\documentclass[12pt,a4paper]{article}

\usepackage{authblk}
\usepackage[square,numbers,sort&compress]{natbib}

\usepackage[utf8]{inputenc}
\usepackage[T1]{fontenc}
\usepackage{mathptmx}       

\usepackage{mathtools}     
\usepackage{amssymb}       
\usepackage{bm}            

\usepackage{graphicx}
\usepackage{float}         
\usepackage{caption}
\usepackage{subcaption}    

\usepackage{dcolumn}       
\usepackage{relsize}       
\usepackage{verbatim}      

\usepackage{algorithm}
\usepackage{algpseudocode} 

\usepackage{xcolor}        
\usepackage{svg}
\usepackage{import}
\usepackage{comment}
\usepackage{soul}          
\usepackage{etoolbox}      
\usepackage{fullpage}      

\makeatletter
\def\@email#1#2{%
 \endgroup
 \patchcmd{\titleblock@produce}
  {\frontmatter@RRAPformat}
  {\frontmatter@RRAPformat{\produce@RRAP{*#1\href{mailto:#2}{#2}}}\frontmatter@RRAPformat}
  {}{}
}%
\makeatother
\begin{document}


\title{Fluid-kinetic multiscale solver for wall-bounded turbulence}


\author[1]{Akshay Chandran}
\author[1]{Praveen Kumar Kolluru}
\author[2]{Berni J. Alder}
\author[3]{Sauro Succi}
\author[1]{Santosh Ansumali}

\affil[1]{Jawaharlal Nehru Centre for Advanced Scientific Research, 
Jakkur, Bengaluru 560064, India}

\affil[2]{Lawrence Livermore National Laboratory, 
Livermore, CA 94550, USA}

\affil[3]{Center for Life Nano Science @ La Sapienza, 
Istituto Italiano di Tecnologia, 00161 Rome, Italy}

\maketitle

\begin{abstract}
We present a two-level (fluid-kinetic) coupling procedure for the
simulation of wall-bounded flows at Reynolds numbers up to thousands.
The method  combines a kinetic Direct Simulation Monte Carlo (DSMC)
treatment of the near-wall layer,  with a high-order Lattice-Boltzmann (HOLB) 
scheme as a fluid solver in the bulk flow. Given the kinetic nature of HOLB,  this coupling is expected to provide a physically 
accurate treatment of the near-wall instabilities which trigger the 
transition to turbulence above a critical threshold around $Re_c \sim 750$.      

The coupled DSMC-HOLB solver is validated by simulating plane Poiseuille 
and Couette flows far from equilibrium,  i.e at finite Knudsen number regimes. 
Based on this validation,  we provide the first preliminary evidence 
that the combination of HOLB and DSMC permits to observe the 
regeneration  cycles of coherent structures which arise 
above a critical value of the Reynolds number. 
This task would be hardly attainable by either of the two solvers 
separately; while DSMC can capture strong near-wall non-equilibrium effects, it lacks 
the compute power to deal with both near-wall and bulk flow at the same time. We look to HOLB to make it computationally feasible to perform such a simulation.
The present two-level coupling procedure
may pave the way to a new generation of fluid-kinetic simulations 
of wall-bounded turbulent flows,  thus helping to gain deeper insights into the role
of wall micro-corrugations in triggering the dynamic instabilities that drive the 
transition to turbulent regimes.  
\end{abstract}


\section{Introduction}

The emergence of fluid turbulence from the underlying molecular dynamics provides 
an adamant example of the amazing complexity that can arise from the collective motion of 
large ensembles of microscopic degrees of freedom. 
For the case of fluid turbulence,  such collective motion results from the
strong spacetime correlations between billions of molecules that behave ``as one".
The extent of such a correlation is dictated by the separation between the relevant 
scales of motion, namely the molecular mean free path versus the shortest turbulent
scales, known as Kolmogorov length, also known as (inner) Knudsen number.

For dilute gases,   described by Boltzmann's kinetic theory,  the 
mean free path is $\lambda  \sim 0.1$ $\mu$m at STP, while  
the Kolmogorov scale depends on the size of the flow and the 
corresponding Reynolds number.
For a moderately turbulent flow in a box 1 meter in size at 
a  Reynolds number $Re \sim 10^4$,  the Kolmogorov   
length is typically $l_k \sim 1000\mu$m, i.e.,  three-four orders
of magnitude above the mean-free path.
A fully resolved turbulent simulation requires mesh-spacings below the Kolmogorov
length, but certainly not thousands of times shorter.  
As a result,  molecular simulations of macroscopic hydrodynamics and  
turbulence are justly viewed as not only computationally unviable but
also physically unnecessary.
This is why, historically,  turbulence has been studied almost exclusively 
at the level of continuum fluid mechanics.  

However, the validity of a continuum treatment of the near-wall flow has been
questioned by a number of authors\textsuperscript{\cite{lockerby2005usefulness,kogan2013rarefied}} ,
mostly on account of the strong non-equilibrium
effects triggered by large gradients in the near-wall region of the flow, which invalidate
the very foundations of the Navier-Stokes equations as a weak-gradient asymptotic limit
of Boltzmann's kinetic theory.  On the other hand,
such strong non-equilibrium effects are naturally incorporated within Boltzmann's
kinetic theory,  hence, at least in principle,  they can be handled by DSMC simulations.

Recent arguments have shown that, due to increased computing power and more advanced
software,  DSMC simulations of turbulence at finite Mach number is a distinct possibility because
the gap between the Kolmogorov length scale and the mean free path shrinks as 
the Mach number is raised, according to the so-called von Karman relation ${\rm Re}={\rm Ma}/{\rm Kn}$.
This relation shows that at a given value of the Reynolds number, increasing the Mach number reflects
in a corresponding increase of the Knudsen number, hence a smaller separation between the
mean free path and the Kolmogorov length.

To study this effect,  DSMC simulations of high Mach number turbulent flows have been conducted
in the recent years\textsuperscript{\cite{pradhan2016transition,gallis2021turbulence}}. 
Although less expensive than molecular dynamics,  a full DSMC simulation of wall-bounded
turbulent flows still remain overly demanding and practically unviable at
Reynolds numbers above a few hundreds.   
For instance, ref. \textsuperscript{\cite{gallis2018gas,gallis2017molecular}} 
simulated the decay to turbulence and minimal Couette flow (MCF) simulations
at Reynolds numbers around,  ${\rm Re} \sim 400$, and a Mach number ${\rm Ma} \sim 0.3$.
These simulations delivered encouraging results but required billions of particles  
and millions of CPU hours on a multi-petascale HPC cluster.
As a result, notwithstanding these remarkable advances, a full-domain DSMC simulation 
at low Mach numbers remains unfeasible for turbulent flows. 

Under such a state of affairs, it becomes natural to seek hybrid strategies, 
coupling DSMC with continuum methods. 
The natural choice for the latter is a discretized solver for the Navier-Stokes equations.
However, since the Navier-Stokes formulation is grounded into the low-Knudsen number approximation
(weak departure from local equilibrium), a formulation closer to kinetic theory would be clearly
preferable. A very appealing alternative in this respect is provided by the  Lattice-Boltzmann (LB) method\textsuperscript{\cite{succi2001lattice,chen1998lattice,benzi1992lattice,succi2018lattice,chen2003extended}}.
Even though historically  LB was limited to low Mach number and low Knudsen 
number flows, in recent years the method has matured to handle high Mach and high Knudsen number flows\textsuperscript{\cite{benzi2006mesoscopic,ansumali2007hydrodynamics,frapolli2015entropic,
namburi2016crystallographic,atif2018higher,montessori2015lattice}}. Since such advances require higher-order stencils as compared to the standard LB schemes, we shall
generically refer to them as to HOLB.

It is therefore natural to investigate what one can gain from coupling DSMC and HOLB solvers. 
Several attempts to couple LB with particle-based methods have been explored
in the recent past\textsuperscript{\cite{montessori2020multiresolution,di2016dsmc}}.
However, they all employ lower-order LB models and a one-way coupling.

\quad In this work, we introduce a two-way coupling between HOLB and DSMC and 
show evidence of the resulting computational gains 
through simulations in the laminar no-slip, transitional and turbulent regimes. 
Wherein the individual solvers are either computationally too expensive or are unable to capture strong non-equilibrium effects near-wall in the aforementioned  regimes, 
the coupled solver,  while still more expensive than a traditional LB solver, 
recovers the correct macroscopic hydrodynamics in regimes that are
inaccessible to standard LB models, at a fraction of the cost of a full-domain DSMC simulation. In the current HOLB, the kinetic nature of the stress dynamics is also reliably modeled.

In the high Knudsen number range, LB suffers from the discrete nature of the model 
and its restricted isotropy, both significantly mitigated by employing high-order stencils.
DSMC\textsuperscript{\cite{bird1994molecular}} remains nevertheless  the method of 
choice in the regime of high Knudsen or high Mach number. 
We can therefore  extend the applicability of  both solvers by combining them in such a way as  to 
use DSMC in regions where non-equilibrium effects are dominant and a HOLB model in the rest of the domain. 

At variance with  initial  works in this area using a DSMC-Lattice Boltzmann\textsuperscript{\cite{di2016dsmc}} and Multi-particle collision dynamics (MPCD)-Lattice Boltzmann\textsuperscript{\cite{montessori2020multiresolution}} coupling in literature which extends LB completely through the domain, we restrict the DSMC solvers exclusively to the molecular regime, namely  only near the wall 
and the HOLB solver only in the bulk, with a handshaking buffer zone for communication between them. 
This gives us the flexibility of moving towards high Knudsen numbers and turbulent flow regimes where the near-wall molecular layer plays a defining  role in capturing boundary non-equilibrium effects. 

This is important to draw the best of both models in a computationally efficient setup. 
In our study,  without loss of any generality, we restrict discussion to the  use of a class of LB 
models known as crystallographic, which use a body-centered cubic (BCC) ordering of the grid points\textsuperscript{\cite{namburi2016crystallographic}}. We have used a 41-velocity crystallographic model, hereby referred to as the RD3Q41 model\textsuperscript{\cite{kolluru2020lattice}}. It was shown that this model ensures higher-order isotropy against the D3Q27 model and is therefore suitable for simulations in the moderate Knudsen number regime with appropriate boundary conditions.  
This model was also chosen due to its ability to describe low Mach number flows and acoustics in a realistic fashion.

\begin{figure}[!htbp]
	\centering
	\captionsetup{justification=centering}
	\includegraphics[scale = .2]{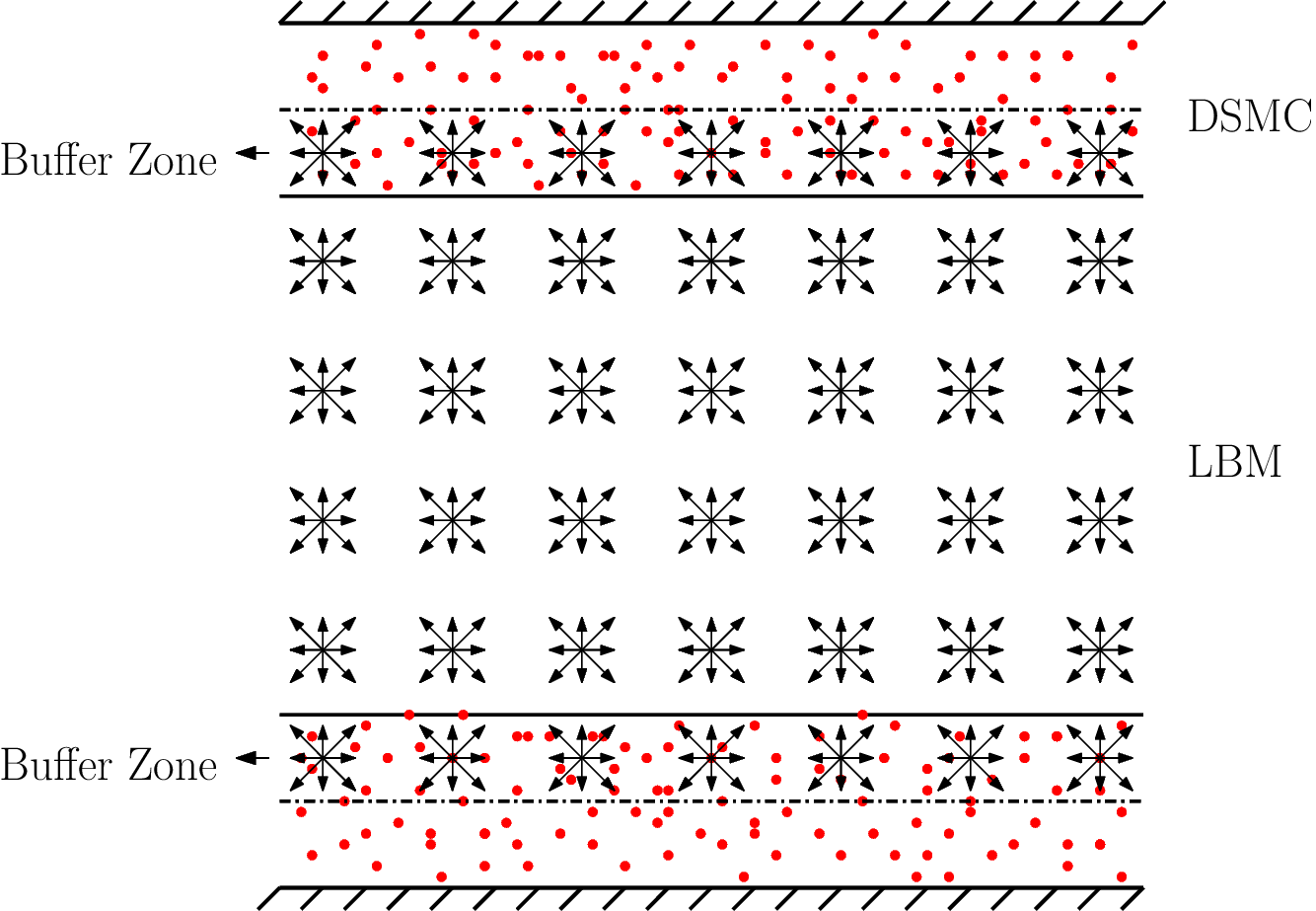}
	\caption{DSMC-HOLB coupling geometry.}
	\label{couplingGeometry}
    \end{figure}

 The schematic in Figure \ref{couplingGeometry} gives a picture of the implemented coupling scheme. The buffer zone represents the region where two-way DSMC-LB and LB-DSMC communication takes place. Any micro-meso-macro coupling scheme faces issues of solving the inverse problem of lifting macroscopic variables to microscopic detailed description and the generation of smooth macroscopic fields from microscopic information. In particular, for the LB to DSMC exchange, the projection of the LB moments such as the density ($\rho$), momentum density ($\rho u_{\alpha}$), stress tensor ($\sigma_{\alpha \beta}$), and the heat flux ($q_{\alpha}$) need to be used to generate the trajectory of  the particles in the DSMC simulations.  Conversely, once particle trajectories are evolved over
 a specified time interval, one needs to generate the discrete distribution to be used in the LB simulation. The communication routine performing this task is described in Section \ref{sec:methods}.

\section{Results}

Finite Knudsen number flow dynamics trigger non-equilibrium effects exposing a dependence on 
higher-order moments beyond those needed for strictly hydrodynamic purposes. 
In this context, one needs to ensure that the dynamics of the higher-order moments is correct at least to the 
leading order, with the most critical higher-order moments being the stress ($\sigma_{\alpha \beta}$), the heat flux ($q_{\alpha}$), and the flux of heat flux ($Q_{\alpha \beta}$). Thus,  any LB scheme meant to match these additional moments  
requires extra discrete velocities\textsuperscript{\cite{ansumali2007hydrodynamics}}. 

Off-grid methods like DSMC and MD do not suffer from this lack of isotropy,  and 
wall effects can be efficiently represented in such schemes. 
However,  as commented earlier on, such a simulation would be quite expensive. 
Thus,  one would hope that with a DSMC layer characterizing the boundary layer, a coupled solver of DSMC paired with 
lattice models with high-order isotropy can dispense from the use of different boundary schemes for flows in the 
finite-Knudsen slip regime.
 
\quad In this section, we present results for the canonical plane Poiseuille and Couette flow cases in the continuum, transitional and turbulent regimes using the multiscale method. The DSMC layer is restricted to a distance of $\approx 4 \lambda$ from either wall beyond which the LB solver covers the rest of the domain with mean temperature $\theta_0 = 0.294896$ corresponding to the reference lattice temperature of the RD3Q41 model\textsuperscript{\cite{kolluru2020lattice}}. For these setups, the Reynolds number is ${\rm Re} = U_c {L}/\nu$, based on the centerline/wall velocity $U_c$ (corresponding to a Mach number of $0.2$), full channel height ${L}$ and kinematic viscosity $\nu$ with nondimensional diffusion time defined as $\overline{t} = t \nu/L^2$.

 \quad In the continuum and transitional regime, the LB region has $40 \times 40 \times 4$ lattice points and 
 the DSMC region has $400 \times 4 \times 40$ cells (Streamwise $\times$ Wall-normal $\times$ Spanwise), with $n_0$ = 100 particles in every cell.  
 With $\Delta y_{\rm LB}/\Delta y_{\rm DSMC} = 10$, we interpolate the moments from the first two LB nodes along the wall-normal direction and use it for particle regeneration in the LB to DSMC exchange step in a region consisting of 100 DSMC cells using Algorithm \ref{algo:particleRegeneration}. 
 
 In the DSMC to LB step in the buffer layer, fluctuations arising from the DSMC region are smoothed out 
 by spatial averaging over a region of $5 \lambda \times 5 \lambda$ in the streamwise and spanwise directions 
 and also over a timeperiod, $t_{\rm avg} \approx 100 \tau$, where $\tau$ is the molecular mean free time.
 
With  128 Million particles in the wall layer, each simulation took approximately 800 CPU hours for the continuum flow and 200 CPU hours for the transitional flow on Intel\textsuperscript{\textregistered} Xeon\textsuperscript{\textregistered} 8,268 processors for   $\bar{t} = 1$. 
 
\begin{figure}[t!p]
\centering
\begin{subfigure}[b]{0.43\textwidth}
    \includegraphics[scale=0.52]{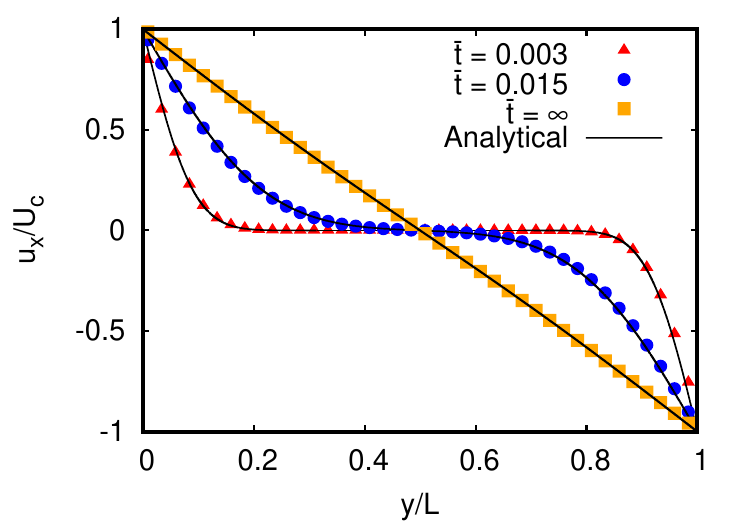}
    \caption{}
     \label{fig:coupledCouette_Re100}
\end{subfigure}
\hspace{20mm}
\begin{subfigure}[b]{0.43\textwidth}
    \includegraphics[scale=0.52]{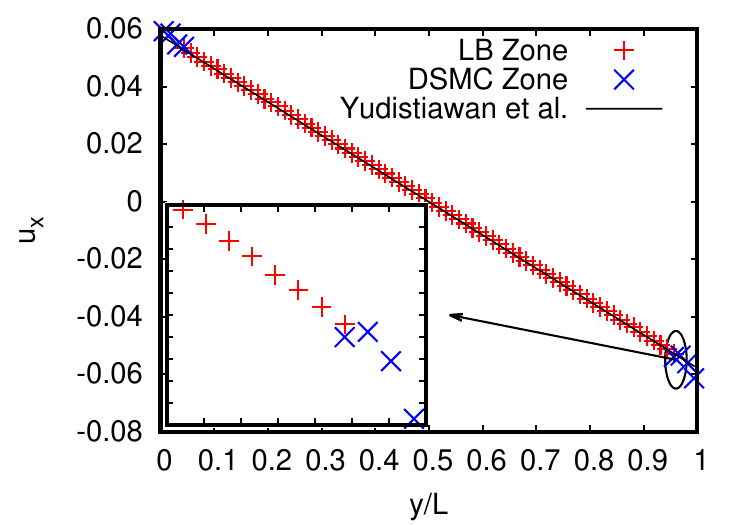}
    \caption{}
    \label{fig:coupledCouette_Knp1}
\end{subfigure}

\begin{subfigure}[b]{0.43\textwidth}
    \includegraphics[scale=0.52]{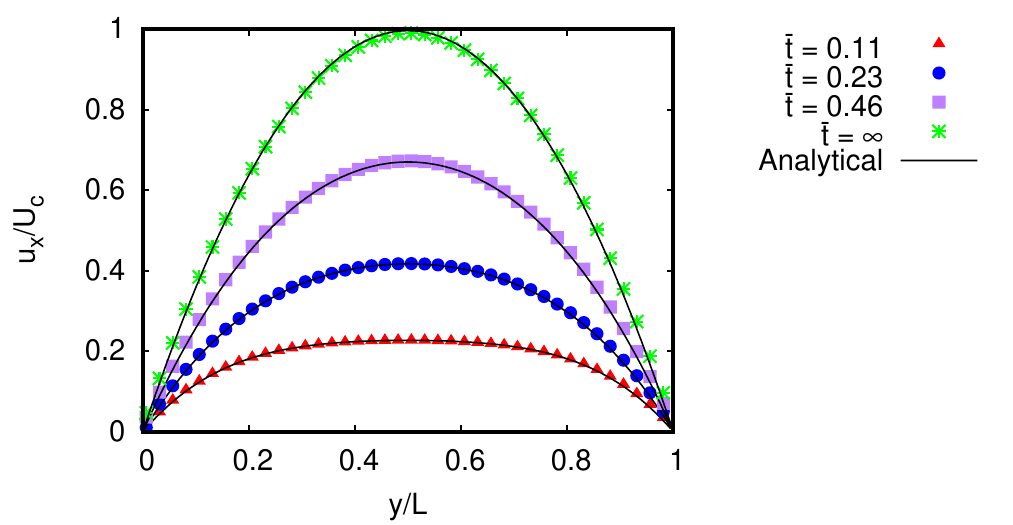}
    \caption{}
    \label{fig:coupledPoiseuille100_ux}
\end{subfigure}
\hspace{20mm}
\begin{subfigure}[b]{0.43\textwidth}
    \includegraphics[scale=0.52]{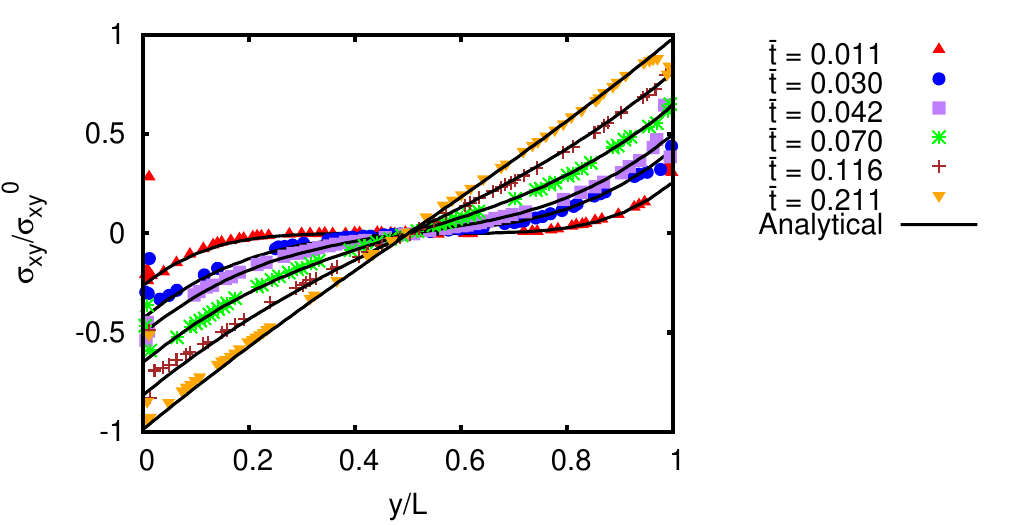}
    \caption{}
    \label{fig:coupledPoiseuille100_stress}
\end{subfigure}
        
\caption{(\protect\subref{fig:coupledCouette_Re100}) Normalized transient velocity profile ($u_x/U_c$, where $U_c$ is the wall velocity) in the continuum regime at various diffusion times compared against analytical results for a planar Couette flow at Re $\approx$ 132, Ma = 0.2. (\protect\subref{fig:coupledCouette_Knp1}) Steady-state velocity ($u_x$) profile for a Couette flow at ${\rm Kn} = 0.1$. A zoomed-in version of the velocity profile at the upper coupling region in the inset. (\protect\subref{fig:coupledPoiseuille100_ux}) Developing stream-wise velocity profile compared with the analytical for a plane Poiseuille flow at Re $\approx$ 132, Ma = 0.2. (\protect\subref{fig:coupledPoiseuille100_stress}) Normalized transient shear stress ($\sigma_{xy}/\sigma_{xy}^0$), where $\sigma_{xy}^0$ is the wall shear stress) profile compared with the analytical solution obtained from a model kinetic equation for the same simulation.}
\label{fig:validationSims}
\end{figure}

To benchmark the code, in Figure \ref{fig:validationSims}\subref{fig:coupledCouette_Re100},  the transient velocity profiles for  simulated  Couette flow are contrasted with  the analytical solution of the Navier-Stokes equations\textsuperscript{\cite{leal2007advanced}}. In order to check the accuracy at finite Knudsen number,  in Figure \ref{fig:validationSims}\subref{fig:coupledCouette_Knp1}, the result from the plane Couette flow simulation at ${\rm Kn} = 0.1$ was contrasted with  the steady-state velocity profile obtained from a model kinetic equation\textsuperscript{\cite{yudistiawan2010higher}}. It is also evident that the overlap regions between LB and DSMC do not show any appreciable discontinuity.

 As a second setup, Figure \ref{fig:validationSims}\subref{fig:coupledPoiseuille100_ux} contrasts the developing stream-wise velocity from our 
 numerical simulation at ${\rm Re} \approx 132$  with the analytical profile  
 for an acceleration driven plane channel flow in the continuum regime at any arbitrary time \textit{t}\textsuperscript{\cite{papanastasiou2021viscous}}. Similarly, Figure \ref{fig:validationSims}\subref{fig:coupledPoiseuille100_stress} also contrasts the transient shear stress profile with the analytical\textsuperscript{\cite{papanastasiou2021viscous}} profile. With the local Mach number $\approx 0.0087$ near the wall, the numerical value of the wall shear stress is also quite small ($\sigma_{xy}^0$ is of $\mathcal{O}(10^{-6})$), leading to significant fluctuations in the higher-order moments near the wall. The standard deviation of these fluctuations decreases with an increase in the number of particles per cell, as detailed in the supplementary material. The beneficial effects of increasing N in taming statistical fluctuations is apparent, although their decay does not seem to obey a uniform $1/\sqrt{\rm N}$ decay in space, which is in line with the non-equilibrium nature of the phenomenon in point.



\begin{figure}[t!p]
\centering
\begin{subfigure}[b]{0.43\textwidth}
    \includegraphics[scale=0.52]{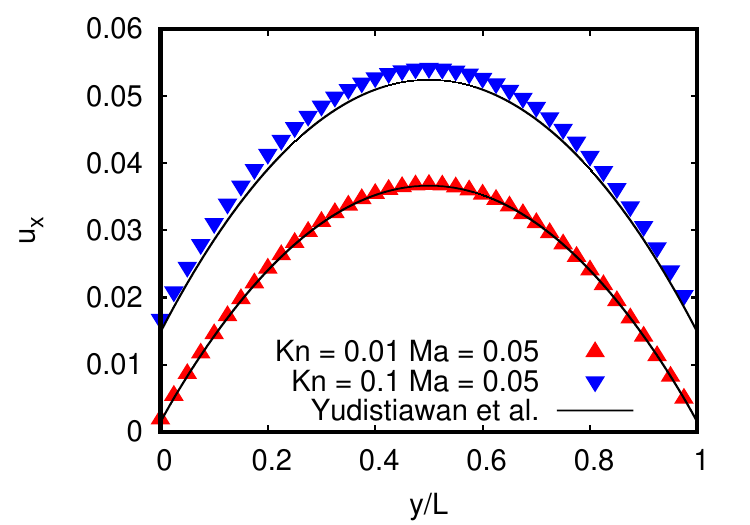}
    \caption{}
     \label{fig:lbCompMap05Kn}
\end{subfigure}
\hspace{20mm}
\begin{subfigure}[b]{0.43\textwidth}
    \includegraphics[scale=0.52]{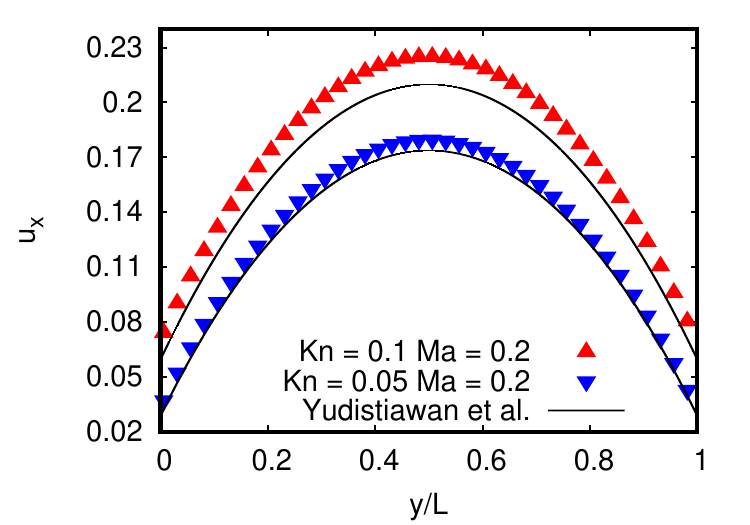}
    \caption{}
    \label{fig:lbCompMap2Kn}
\end{subfigure}

\caption{Steady-state velocity ($u_x$) profiles for a Poiseuille flow using an LBM RD3Q41 model\textsuperscript{\cite{kolluru2020lattice}} at (\protect\subref{fig:lbCompMap05Kn}) ${\rm Ma} = 0.05$ and ${\rm Kn} = 0.01,0.1$, (\protect\subref{fig:lbCompMap2Kn}) ${\rm Ma} = 0.2$ and ${\rm Kn} = 0.05,0.1$.}
\label{fig:transitionalLB_poiseuille}
\end{figure}

\begin{figure}[t!p]
\centering
\begin{subfigure}[b]{0.43\textwidth}
    \includegraphics[scale=0.52]{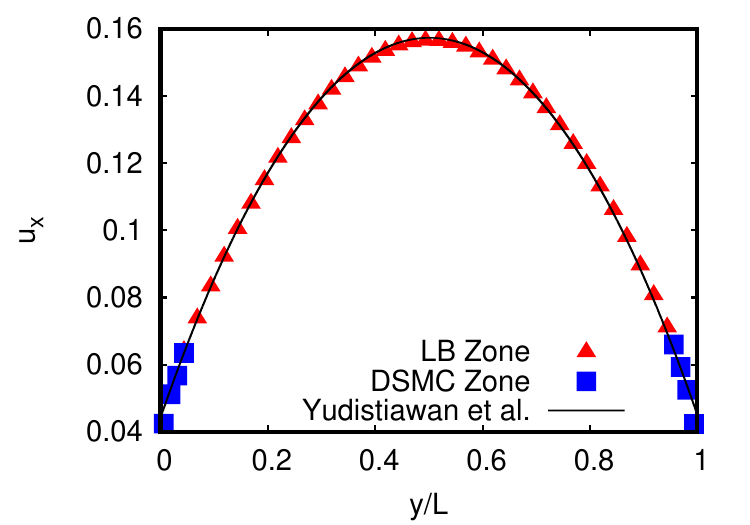}
    \caption{}
    \label{fig:knp1poiseuille_ux}
\end{subfigure}
\hspace{20mm}
\begin{subfigure}[b]{0.43\textwidth}
    \includegraphics[scale=0.52]{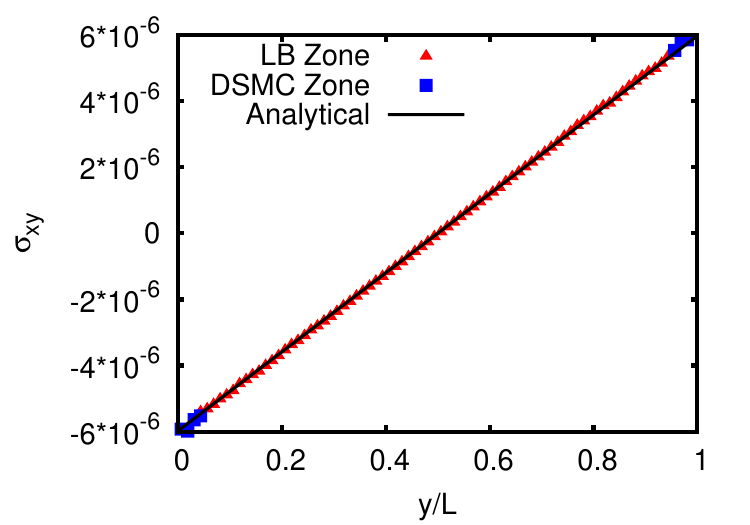}
    \caption{}
    \label{fig:knp1poiseuille_stress}
\end{subfigure}

\begin{subfigure}[b]{0.43\textwidth}
    \includegraphics[scale=0.52]{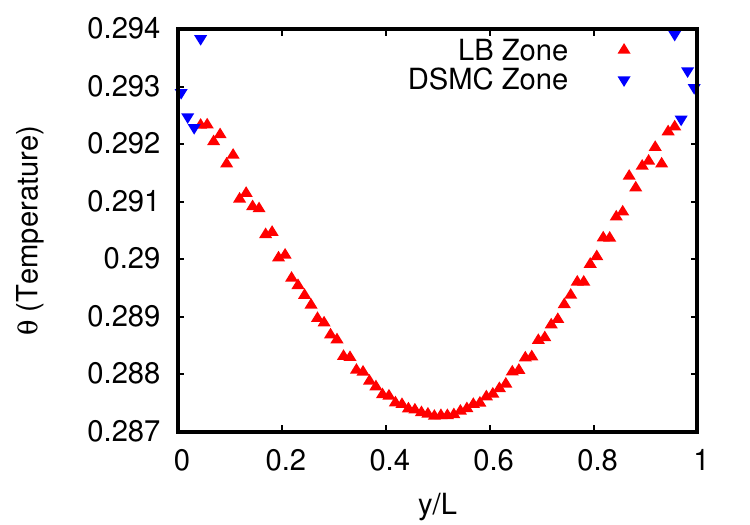}
    \caption{}
    \label{fig:knp1stressTensor_theta}
\end{subfigure}
\hspace{20mm}
\begin{subfigure}[b]{0.43\textwidth}
    \includegraphics[scale=0.52]{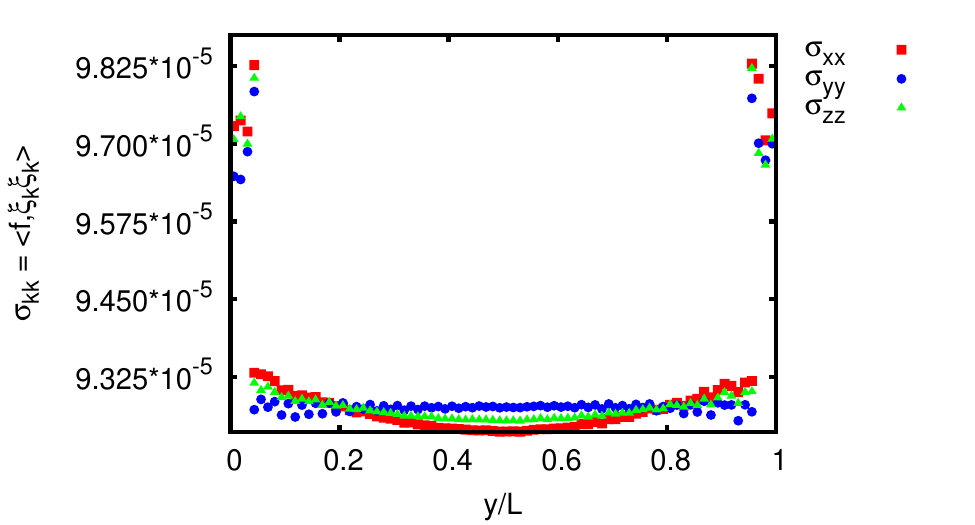}
    \caption{}
    \label{fig:knp1stressTensor_sigmaDiag}
\end{subfigure}

\caption{Steady-state (\protect\subref{fig:knp1poiseuille_ux}) velocity ($u_x$) and (\protect\subref{fig:knp1poiseuille_stress}) Shear stress ($\sigma_{xy}$) profiles for a Poiseuille flow at ${\rm Kn} = 0.1$. Steady-state (\protect\subref{fig:knp1stressTensor_theta}) temperature ($\theta$) and (\protect\subref{fig:knp1stressTensor_sigmaDiag}) stress tensor components ($\sigma_{xx}, \sigma_{yy}, \sigma_{zz}$) profiles for a poiseuille flow at ${\rm Kn} = 0.1$.}
\label{fig:transitionalCoupled_poiseuille}
\end{figure}

Limitations arising at finite Knudsen numbers in the lattice boltzmann method can be seen in Figures \ref{fig:transitionalLB_poiseuille}\subref{fig:lbCompMap05Kn} \& \ref{fig:transitionalLB_poiseuille}\subref{fig:lbCompMap2Kn}, wherein at progressively higher Mach and Knudsen numbers, the deviations from ref.\cite{yudistiawan2010higher} show an increase. Using the coupled scheme, Figure \ref{fig:transitionalCoupled_poiseuille}\subref{fig:knp1poiseuille_ux} \& \ref{fig:transitionalCoupled_poiseuille}\subref{fig:knp1poiseuille_stress} compares the steady-state velocity and shear stress profiles against their analytical solutions at ${\rm Kn} = 0.1$. Here the higher-order moment, $\sigma_{xy}$ shows a good match with the numerically approximate value.  Higher-order moments ($\sigma_{xx}, \sigma_{yy}, \sigma_{zz}$) as shown in Figure \ref{fig:transitionalCoupled_poiseuille}\subref{fig:knp1stressTensor_sigmaDiag} show a smooth transition in the coupling region, with deviations in the DSMC region against the bulk arising at $\mathcal{O}(10^{-6})$. Meanwhile, in Figure \ref{fig:transitionalCoupled_poiseuille}\subref{fig:knp1stressTensor_theta} the temperature profile shows the relative cooling of the bulk as compared to the walls.

\subsection{Turbulent plane Couette flow}

A turbulent simulation was performed for a Minimal Couette flow (MCF) 
setup with the bulk LB region consisting of 
$720 \times 200 \times 400$ nodes and each DSMC region consisting of
 $720 \times 4 \times 1200$ cells and $n_0 = 200$ particles per cell 
 (approximate  memory requirement of 100 GB for the DSMC regions with about 1.4 billion particles). 
 The solver took 120 hours using 2400 cores on Intel\textsuperscript{\textregistered} Xeon\textsuperscript{\textregistered} 8268 processors to reach 30 convection times ($L/U_c$). Beyond which averaging was done over another 10 convection times.

\quad At the fully developed state, the velocity profile in plane Couette flow is monotonic in the laminar and turbulent regimes. 
While linear stability theory predicts that two-dimensional disturbances are damped for plane Couette and pipe flows\textsuperscript{\cite{wasow1953small,gallagher1962behaviour}}, experiments have shown that 
for large enough initial disturbances, the flow transitions to turbulence subcritically\textsuperscript{\cite{davies1928experimental,kao1970experimental,
patel1969some}}. Experimental\textsuperscript{\cite{tillmark1992experiments}} and numerical\textsuperscript{\cite{lundbladh1991direct}} studies performed to determine the transitional ${\rm Re}$ have shown excellent agreement, with the former 
predicting the critical threshold at $720 \pm 20$, while the latter at $750$. 

When the initial flow field is disturbed using a finite-amplitude perturbation, 
plane Couette flow transitions at ${\rm Re} \approx 750$ .  
Thousands of experimental and  numerical studies performed on pipe 
flow have shown that the turbulent state in such linearly stable shear flows belongs to a transient 
turbulent chaotic saddle\textsuperscript{\cite{hof2006finite}}. 
The implication of this is that rather than a critical ${\rm Re}$ beyond which turbulence is 
believed to self-sustain indefinitely,  there exists a time scale $\tau({\rm Re})$ defining 
the lifetime of a turbulent state.  More precisely,  the probability distribution of lifetimes increases 
exponentially with the Reynolds number but stays finite,  meaning that 
with long enough observation times, all linearly stable shear flows are seen 
to decay and relaminarize \textsuperscript{\cite{hof2006finite}}. 
Similar observations apply to plane Couette flows\textsuperscript{\cite{schmiegel1997fractal}}. 

For a Couette flow, following an exponential scaling of lifetime with ${\rm Re}$, we get a 
lifetime of $\mathcal{O}(10^{24})$ convection times at ${\rm Re} = 1300$,  far longer than with any
foreseeable numerical and experimental capabilities.


\begin{figure}[!htpb]
\centering
\begin{subfigure}[b]{0.5\textwidth}
    \includegraphics[scale=0.52]{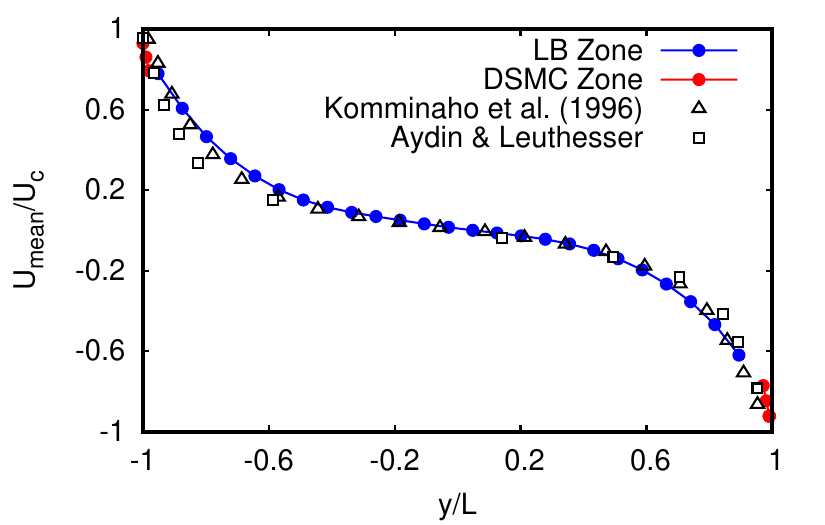}
    \caption{}
    \label{fig:velocityCouplingComp_meanVel}
\end{subfigure}

\begin{subfigure}[b]{0.43\textwidth}
    \includegraphics[scale=0.52]{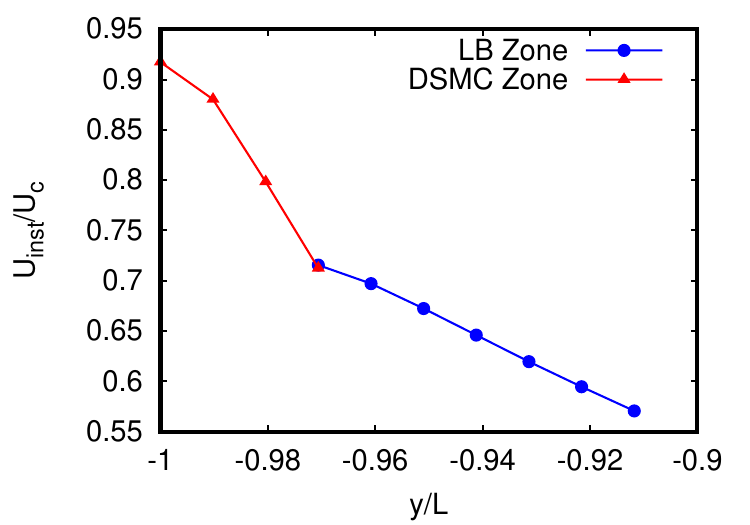}
    \caption{}
    \label{fig:velocityCouplingComp_interfaceLower_instVel}
\end{subfigure}
\hspace{20mm}
\begin{subfigure}[b]{0.43\textwidth}
    \includegraphics[scale=0.52]{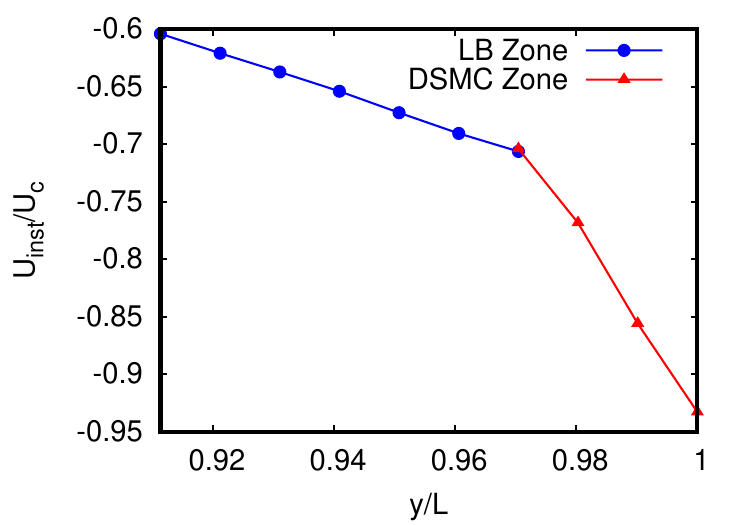}
    \caption{}
    \label{fig:velocityCouplingComp_interfaceUpper_instVel}
\end{subfigure}

\caption{(\protect\subref{fig:velocityCouplingComp_meanVel}) Turbulent steady-state mean velocity profile from the coupled solver compared against the numerical profile from ref.\cite{komminaho1996very} and experimental profile from ref.\cite{aydin1979novel}. Here, $U_c$ represents the imparted wall velocity. (\protect\subref{fig:velocityCouplingComp_interfaceLower_instVel}) \& (\protect\subref{fig:velocityCouplingComp_interfaceUpper_instVel}) A zoomed-in version of the lower (left) and upper (right) coupling interface in the instantaneous stream-wise velocity profile.}
\label{fig:velocityCouplingComp}
\end{figure}

\quad The influence of thermal fluctuations on the deterministic Navier-Stokes equations, specifically on the dissipation range spectrum was first explored experimentally by Betchov\textsuperscript{\cite{betchov1957fine}}. Recent investigations\textsuperscript{\cite{bandak2022dissipation}} using shell models of turbulence show that these thermal fluctuations have significant impact up until the Kolmogorov scale, where most often a continuum description is assumed. Additionally, their study also reveals
the propagation of fluctuations into the dissipation range due to inertial-range intermittency. Molecular simulations of homogenous isotropic turbulence further confirm the existence of these fluctuating modes in the dissipation range\textsuperscript{\cite{mcmullen2022navier,bell2022thermal}}. While continuum NS simulations predict an exponential decay of the sub-kolmogorov energy spectra, these simulations show a quadratic growth in this regime.

\quad In wall-bounded turbulent flows, these fluctuations can be significant in the viscous sublayer. Using a coupled solver with the DSMC layer provides a suitable 
and alternative boundary condition to the micro-roughness of the wall,  and consequently,
it can help in gaining deeper insights into this fundamental problem. 
Similarly, the transition Reynolds number for highly compressible flows is also unknown, hence
the present simulations of turbulent plane Couette flow provide an initial step  to explore this 
important question.

\quad Turbulent flow simulations of MCF at Reynolds number of ${\rm Re} \approx 1318$ were performed using the present
coupled method. Similar simulations with standard  LB solver relaminarizes after a few convection times. 
The coupled solver in this case helps in capturing the slow regeneration cycles in turbulent MCF. 
The domain was initially disturbed using a finite-amplitude perturbation generated using a curl-based Perlin noise\textsuperscript{\cite{bridson2007curl}}.

\begin{figure}[!htpb]
\centering
\begin{subfigure}[b]{\textwidth}
    \includegraphics[scale=0.2]{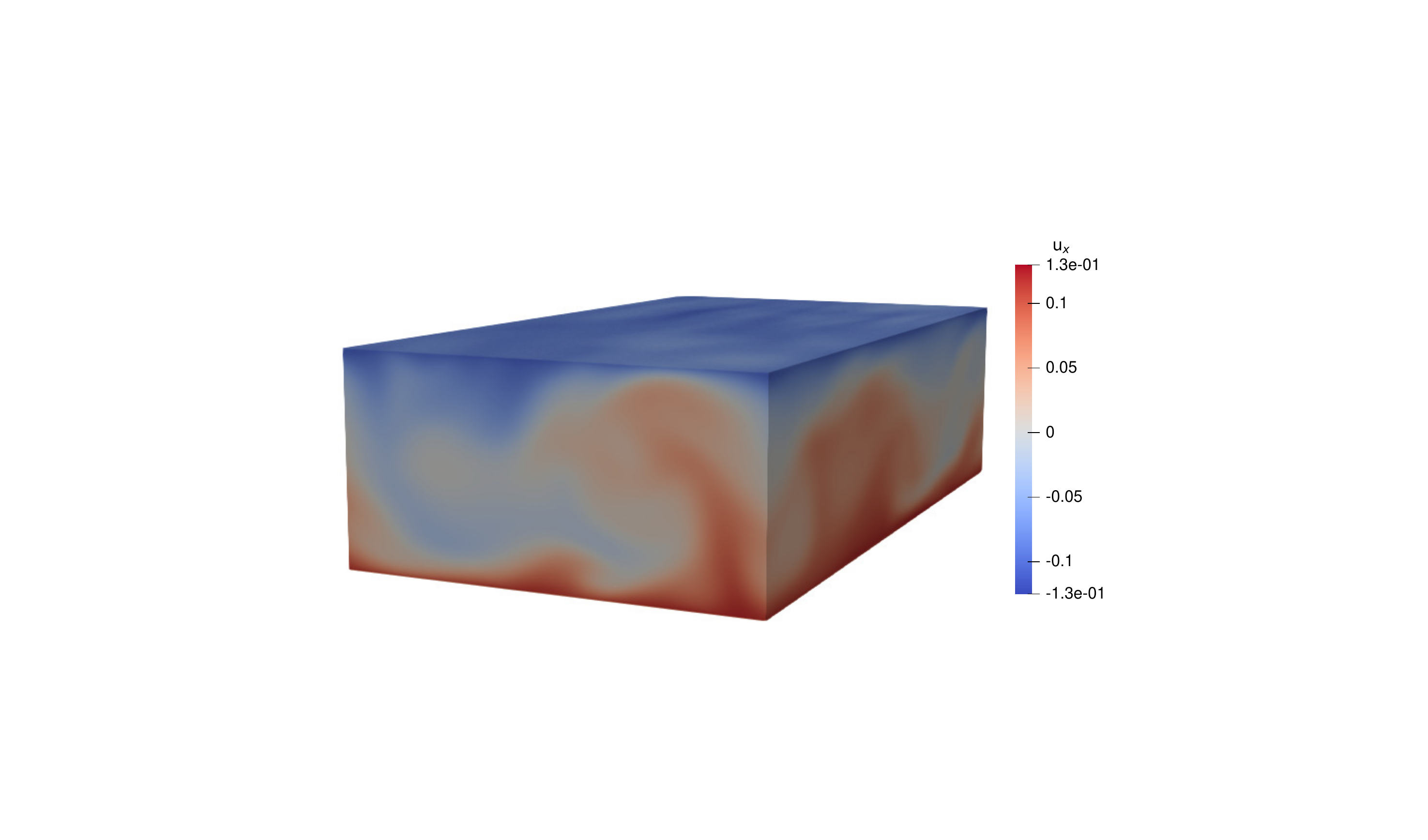}
    \includegraphics[scale=0.2]{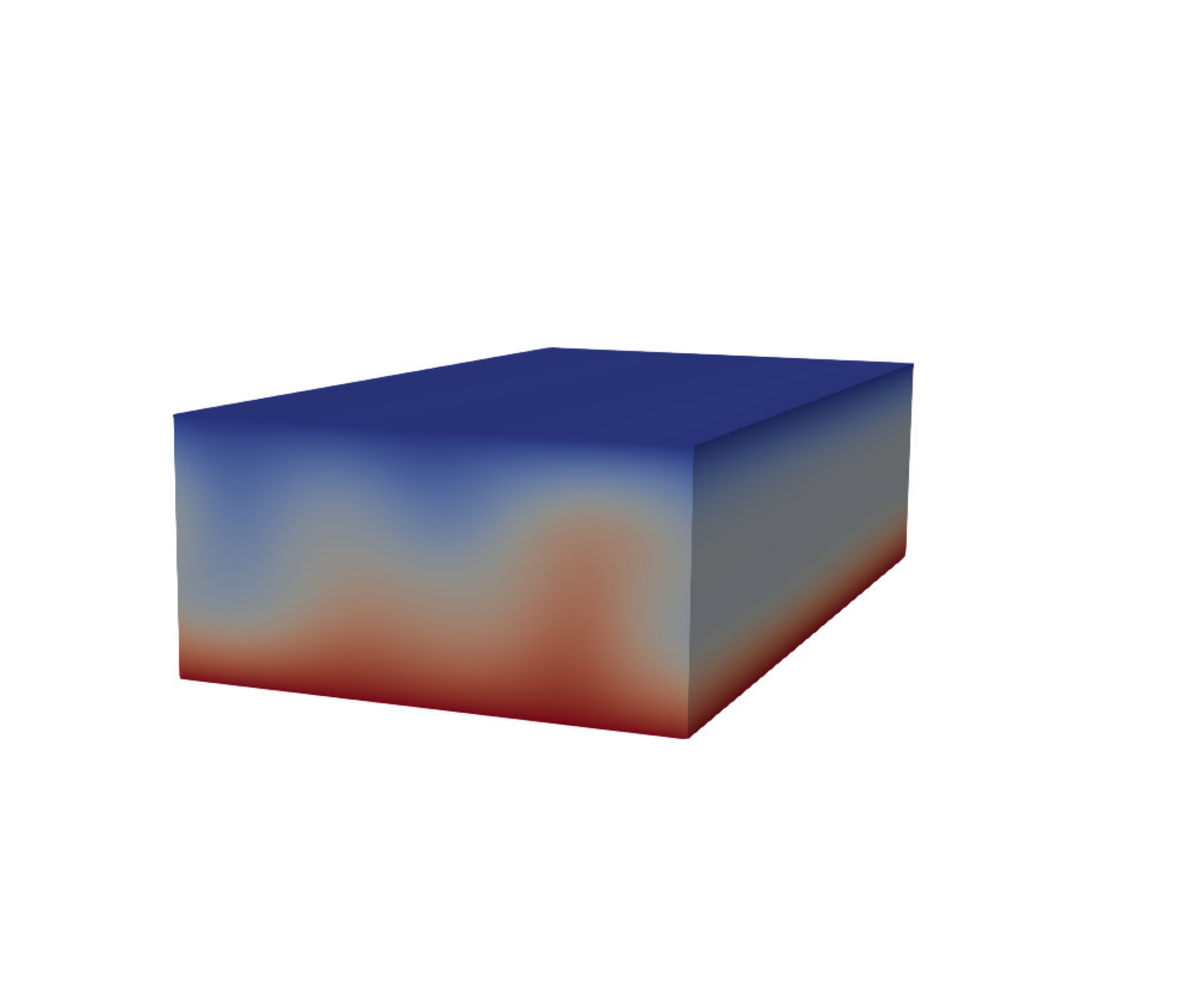}
    \caption{}
    \label{fig:mcfVelocityField}
\end{subfigure}

\begin{subfigure}[b]{0.24\textwidth}
    \includegraphics[width=\linewidth,scale=0.07]{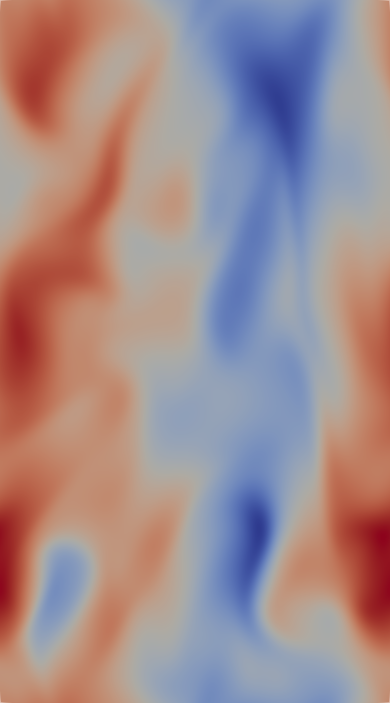}
    \caption{}
    \label{fig:mcfCoherentStructures_r2_2000}
\end{subfigure}
\begin{subfigure}[b]{0.24\textwidth}
    \includegraphics[width=\linewidth,scale=0.07]{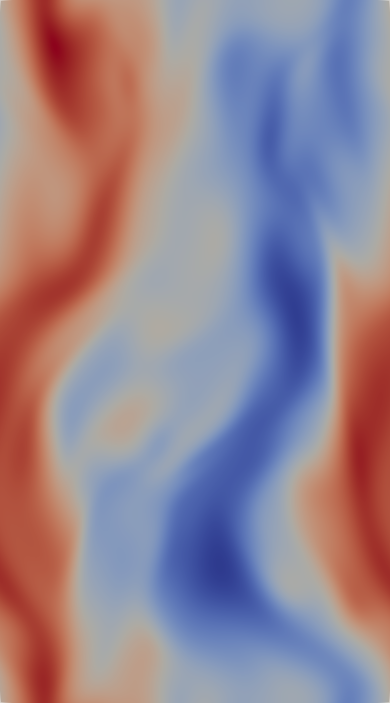}
    \caption{}
    \label{fig:mcfCoherentStructures_r3_2000}
\end{subfigure}
\begin{subfigure}[b]{0.24\textwidth}
    \includegraphics[width=\linewidth,scale=0.07]{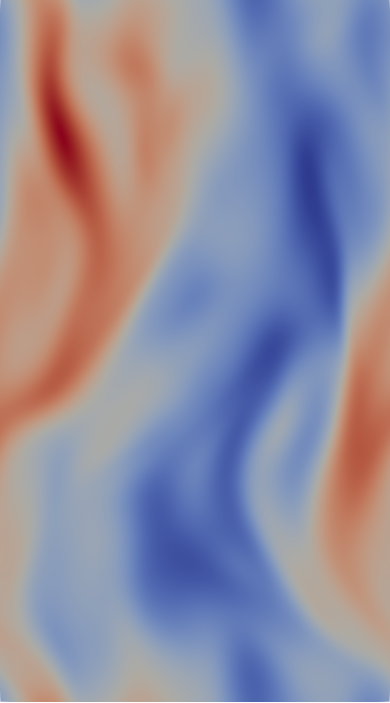}
    \caption{}
    \label{fig:mcfCoherentStructures_r4_500}
\end{subfigure}
\begin{subfigure}[b]{0.24\textwidth}
    \includegraphics[width=\linewidth,scale=0.07]{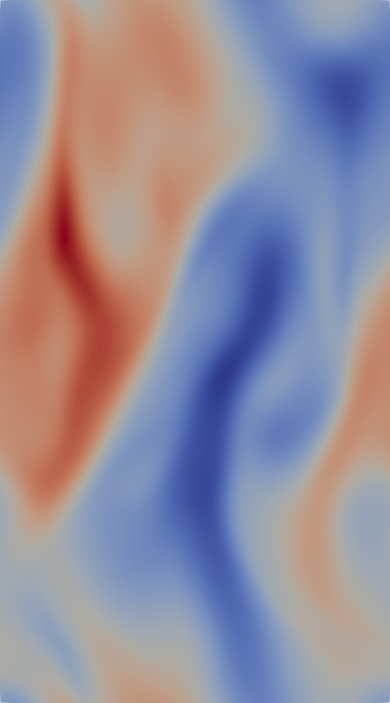}
    \caption{}
    \label{fig:mcfCoherentStructures_r4_2200}
\end{subfigure}

\caption{(\protect\subref{fig:mcfVelocityField}) Streamwise velocity ($u_x$) field for a DSMC-LB coupled Minimal Turbulent Couette flow (taken after 20 convection times) compared against an LB simulation with the same parameters. (\protect\subref{fig:mcfCoherentStructures_r2_2000})-(\protect\subref{fig:mcfCoherentStructures_r4_2200}) Cycle of Regeneration and decay of coherent structures in Minimal Couette flow at the channel mid-plane. Snapshots were taken at 14, 20, 22, and 25 convection times respectively.}
\label{fig:mcfVelField_CoherentStructures}
\end{figure}

\quad Figure \ref{fig:velocityCouplingComp}\subref{fig:velocityCouplingComp_meanVel} compares the mean velocity obtained after spatial averaging along the span-wise and stream-wise directions as well as time averaging over 10 convection times. The numerical profile from ref.\cite{komminaho1996very} have performed an averaging over 620 convection times. As there is constant decay and regeneration in the near-wall region in turbulent Couette flow, an averaging over a timescale larger than the regeneration time scale would be required. We still observe a good agreement between the DSMC-LB coupled solver and the profiles from ref.\cite{komminaho1996very} and ref.\cite{aydin1979novel} near the channel mid-plane. The deviations near the wall result due to the noise from the DSMC solver in the region. We performed similar simulations using $n_0$ = 30 to 100 and observed the averaged velocity profile to deviate further as we lower the number of particles per cell. In Figures \ref{fig:velocityCouplingComp}\subref{fig:velocityCouplingComp_interfaceLower_instVel} \& \ref{fig:velocityCouplingComp}\subref{fig:velocityCouplingComp_interfaceUpper_instVel}, we show the coupling interface using the instantaneous stream-wise velocity profile.

Figure \ref{fig:mcfVelField_CoherentStructures}\subref{fig:mcfVelocityField} presents a visualization of the streamwise velocity ($u_x$) field taken after 20 convection times and is compared against a corresponding LB simulation run using the same parameters. One can see the presence of coherent structures in both simulations. 
The LB velocity profile is smoother as compared to its DSMC-HOLB counterpart and results 
in the HOLB simulation to relaminarize after about fifty convection times. 
For linearly stable flows, finite lifetimes have been observed in similar studies on pipe flows\textsuperscript{\cite{hof2006finite}}. A further perturbation to the relaminarized state leads to the cycles of regeneration and decay. As the DSMC layer injects thermal noise into the bulk LB, such a relaminarization could take longer to achieve.

Figure \ref{fig:mcfVelField_CoherentStructures}\subref{fig:mcfCoherentStructures_r2_2000}-\subref{fig:mcfCoherentStructures_r4_2200} shows the streamwise velocity contours at the channel mid-plane taken at various convection times. We see long coherent structures that undergo cycles of regeneration and decay indicating the sustenance of turbulence using the coupled solver. As observed in similar studies\textsuperscript{\cite{gallis2018gas}}, we observe two vortices that occupy half the spanwise width with anti-parallel orientation along the streamwise direction. 
Where  ref. \cite{gallis2018gas} performed full-domain DSMC simulations using almost 600 million CPU hours, our coupled solver has achieved a similar feat, yet on a smaller scale, using 0.3 million CPU hours.  
Although full-domain LB simulations take significantly lower time, to study the effect of sustained turbulence, relaminarization in LB simulations are a hurdle, as one needs to provide frequent disturbances to the LB simulation in order
to sustain turbulence.  Therefore,  the coupled DSMC-LB solver can be used to study the effects of turbulence in linearly stable shear flows,
at an affordable computational cost while  providing accuracy comparable with microscopic simulations\textsuperscript{\cite{gallis2018gas}}.

\section{Discussion}

In this work, a kinetic-continuum approach using DSMC and high-order 
LB\textsuperscript{\cite{kolluru2020lattice}} was developed. 

The coupling was tested in the continuum, transitional and turbulent regimes to validate the coupled solver. 
All simulations were performed using the channel flow geometry with a DSMC wall layer and an 
LB region in the bulk. 
In the continuum regime, test cases were performed on the canonical plane Couette and Poiseuille 
flows at ${\rm Re} = \mathcal{O}(10^2)$ and ${\rm Ma} = 0.2$ and the corresponding 
results served as a benchmark to show the ability and efficiency of the coupled solver in 
reproducing Navier-Stokes hydrodynamics with higher-order moments in good agreement with theory.  

A set of simulations were then performed on canonical cases in the transitional regime to demonstrate the smooth transition of moments 
across the coupling region, with deviations arising only at $\mathcal{O}(10^{-6})$ after sufficient
statistical averaging. 

Finally,  first-time turbulent flow simulations using the coupled solver were performed for the 
Minimal Turbulent Couette flow at ${\rm Re} \approx 1318$ and ${\rm Ma} = 0.2$. 
We observe the decay and regeneration cycles at small time scales and the characteristic 
S-shape of the flow is reproduced and validated against experimental and numerical simulations. 
Future work using the coupled solver will focus on assessing transition thresholds in turbulent plane Poiseuille flows. 
With the inherent thermal noise from the DSMC wall layer, artificial forcing to transition Poiseuille 
flows may become redundant, and work along these lines is currently in progress.
Other areas of focus are systems with high thermal gradients where local Mach numbers exceed the standard LB limits.

\quad To summarize, our results point towards Bernie Alder’s early hunch, namely that 
the revival of turbulence cannot be modeled by the Navier-Stokes equations because continuum 
fluid mechanics fails to account for the strong non-equilibrium effects that arise in the near-wall region. 
A similar statement would also apply to standard hydrodynamic LB methods.
High-order LB are better positioned because they can accommodate 
mild non-equilibrium effects, but still, this is not sufficient for quantitative 
purposes, unless they are coupled to a DSMC treatment of the boundary layer. 
An appealing  possibility for the future is to replace DSMC with a stochastic near-wall 
forcing term in the HOLB formulation.  However, the multiscale LB-DSMC procedure presented in this work 
is physically more sound and arguably also more robust from the numerical standpoint.

\section{Methods \label{sec:methods}}

\textbf{Particle regeneration algorithm.} Particle generation follows a Monte-Carlo sampling procedure akin to ref.\cite{garcia1998generation} .

\begin{algorithm}[H]
\caption{Particle Regeneration}
\label{algo:particleRegeneration}
\begin{algorithmic}[1]
\While{\textit{N$_i$} $\leq$ \textit{N$_{\rm bufferCells}$}}
\State \textit{n$_{\rm poisson}$} = poissonDist(\textit{$\rho_{\rm LB}V_c$})
\While{$n_0$ $\leq$ \textit{n$_{\rm poisson}$}  }
\State Compute maxima of the coefficients of the Hermite polynomials
\begin{equation*}
C = \rm max(|a_{\rm LB}^{(0)}|,|a_{{\rm LB},\alpha}^{(1)}|,|a_{{\rm LB},\beta \gamma}^{(2)}|,|a_{{\rm LB},\kappa}^{(3)}|)
\end{equation*}
\State Generate\textsuperscript{\cite{agrawal2018molecular}} random particle velocities $\boldsymbol{\xi}$ from the Maxwell-Boltzmann distribution with temperature $\theta_{\rm LB}$
\State Select $\boldsymbol{\xi}$ \textbf{if} \textit{C$\mathcal{R}$ $\leq$ f($\boldsymbol{ x, \xi}$, t)}, where $\mathcal{R}$ $\epsilon$ (0,1); \textbf{else} \textbf{goto} step 5
\State Compute particle velocity using \textit{c$_{\alpha}$} = \textit{$\xi_{\alpha}$} + \textit{u$_{{\rm LB},\alpha}$}
\EndWhile
\EndWhile

\State rescale all \textit{c$_{\alpha}$} within cell to match $u_{{\rm LB},\alpha}$ and $\theta_{\rm LB}$
\end{algorithmic}
\end{algorithm}

\textbf{DSMC-LBM coupling scheme.} The first ingredient to set up the coupling between the two methods is establishing
an information-exchange protocol between the two.
This is naturally provided by Grad's moment method,  which consists of a systematic expansion
of the Boltzmann distribution onto a suitable set of basis functions in velocity space, typically tensor Hermite
polynomials.  In other words, Grad's moment method provide a natural way to lift extended list of macroscopic variables to microscopic description of distribution functions\textsuperscript{\cite{grad1949kinetic}}. 

The  distribution function $f({\bm x},{\bm \xi},t)$ is expanded in terms of  Hermite orthonormal polynomial  of the peculiar velocity ${\bm \xi} = {\bm c}$ - ${\bm u}$  as
\begin{equation}
f({\bm x},{\bm \xi},t) = w({\bm \xi}) \sum_{n=0}^{N} {\bm a}^{(n)}({\bm x},t) \mathcal{H}^{(n)}({\bm \xi}),
\label{eqn:gradGenExp}
\end{equation}
where order $N$ is  pre-specified and $\mathcal{H}^{(n)}({\bm \xi})$ are the Hermite polynomials 
at $n^{th}$ order, ${\bm a}^{(n)}({\bm x},t)$ are their corresponding coefficients, $w({\bm \xi})$ is the  
Maxwell-Boltzmann distribution at reference condition for the various Hermite polynomials\textsuperscript{\cite{grad1949kinetic}}. 

We achieve macro-micro communication  through the generation of a $13$-moments Grad-distribution function  
in explicit form as follows:\textsuperscript{\cite{grad1949kinetic}} 
\begin{equation}
f({\bm x},{\bm \xi},t) = F^{\rm grad}({\bm x},{\bm \xi},t) = w({\bm \xi}) \Biggl[\rho + \frac{\sigma_{\beta \gamma}}{2 \theta_0^2} (\xi_{\beta} \xi_{\gamma} - \theta_0 \delta_{\beta \gamma}) + \frac{q_{\kappa}}{5 \theta_0^3} (\xi^2 \xi_{\kappa} - 5 \theta_0 \xi_{\kappa})\Biggr].
\label{eqn:gradDistFuncCont}
\end{equation}
Once Grad's  distribution function is formed,  particle positions and velocities can 
be generated using a Monte-Carlo sampling.  The procedure for converting from this non-equilibrium distribution 
to the individual particle velocities is outlined in Algorithm \ref{algo:particleRegeneration}.

In the LB framework, one uses  the discrete form of the Hermite expansion
\textsuperscript{\cite{shan1998discretization}} where  the weights ($w_i$) associated 
with the discrete velocity set are used in place of the discrete equilibrium function. 
Through a similar process of Gram-Schmidt orthonormalization and discrete orthonormality 
condition, we obtain the discrete Hermite polynomials and their coefficients 
\textsuperscript{\cite{chikatamarla2006grad}}. 
Using these coefficients,  one  writes down the discrete Grad-distribution function as a weighted 
sum of discrete Hermite polynomials:

\begin{equation}
f_i = w_i \Biggl[\rho + \frac{j_{\alpha} c_{i \alpha}}{\theta_0} + \frac{(P_{\beta \gamma} - \rho \theta_0 \delta_{\beta \gamma})}{2 \theta_0^2} (c_{i\beta} c_{i \gamma} - \theta_0 \delta_{\beta \gamma}) + \frac{(Q_{\kappa} - 5 j_{\kappa} \theta_0)}{10 \theta_0^3} (c_i^2 c_{i \kappa} - 5 \theta_0 c_{i \kappa})\Biggr].
\label{eqn:gradDistFuncDiscrete}
\end{equation}

\quad For the DSMC to LB coupling, the moments from the DSMC region are spatially averaged over a region extending far 
beyond the DSMC cell dimension. This is to ensure that the discrete equilibrium in the corresponding LB nodes 
in the buffer layer remain positive.  Here, the moments ($\rho, j_{\alpha}, P_{\alpha \beta}, Q_{\kappa}$) in Equation (\ref{eqn:gradDistFuncDiscrete}) are spatially and temporally averaged from the DSMC cells in the buffer layer.

The steps followed for a single iteration of the coupling algorithm are as follows:
\begin{enumerate}
\item DSMC advection and collision for \textit{n} ($= \Delta t_{\rm LB}/\Delta t_{\rm DSMC})$ steps.
\item DSMC to LB moment exchange in the buffer layer through the discrete Grad distribution function (Equation (\ref{eqn:gradDistFuncDiscrete})) followed by LB streaming and collision.
\item LB to DSMC moment exchange in the buffer layer with particle velocity regeneration using a Monte-Carlo sampling procedure (Algorithm \ref{algo:particleRegeneration}).
\end{enumerate}

Algorithm \ref{algo:couplingAlgorithm} provides a pseudo code for clarity.

\begin{algorithm}
\caption{Coupling methodology}
\label{algo:couplingAlgorithm}
\begin{algorithmic}[1]
\For{\textit{i,j} $\epsilon$ DSMC region}
\While{\textit{steps} \textless  \textit{n}}
\State advection via equations of motion
\State particle collisions
\State calculate moments in buffer layer
\EndWhile
\EndFor

\State calculate LB populations through Equation (\ref{eqn:gradDistFuncDiscrete}) in the buffer layer

\For{\textit{i,j} $\epsilon$ LB region}
\State advect
\State collide
\State calculate moments in buffer layer
\EndFor

\State interpolate LB moments and regenerate particles using Algorithm \ref{algo:particleRegeneration}
\State \textbf{goto} \emph{top}
\end{algorithmic}
\end{algorithm}


\medskip
\textbf{Acknowledgements} \par 
The support and the resources provided by the ‘PARAM Yukti Facility’ under the National Supercomputing Mission, Government of India at the Jawaharlal Nehru Centre For Advanced Scientific Research are gratefully acknowledged. One of the authors (SS) kindly acknowledges funding from the European Research Council under the Horizon 2020 Programme Grant Agreement n. 739964 ("COPMAT").

\bibliographystyle{unsrt}
\bibliography{References}

@article{bandak2022dissipation,
  title={Dissipation-range fluid turbulence and thermal noise},
  author={Bandak, Dmytro and Goldenfeld, Nigel and Mailybaev, Alexei A and Eyink, Gregory},
  journal={Physical Review E},
  volume={105},
  number={6},
  pages={065113},
  year={2022},
  publisher={APS}
}

@article{bell2022thermal,
  title={Thermal fluctuations in the dissipation range of homogeneous isotropic turbulence},
  author={Bell, John B and Nonaka, Andrew and Garcia, Alejandro L and Eyink, Gregory},
  journal={Journal of fluid mechanics},
  volume={939},
  pages={A12},
  year={2022},
  publisher={Cambridge University Press}
}

@article{betchov1957fine,
  title={On the fine structure of turbulent flows},
  author={Betchov, Robert},
  journal={Journal of Fluid Mechanics},
  volume={3},
  number={2},
  pages={205--216},
  year={1957},
  publisher={Cambridge University Press}
}

@article{bridson2007curl,
  title={Curl-noise for procedural fluid flow},
  author={Bridson, Robert and Houriham, Jim and Nordenstam, Marcus},
  journal={ACM Transactions on Graphics (ToG)},
  volume={26},
  number={3},
  pages={46--es},
  year={2007},
  publisher={ACM New York, NY, USA}
}

@article{gallis2018gas,
  title={Gas-kinetic simulation of sustained turbulence in minimal Couette flow},
  author={Gallis, Michail A and Torczynski, John R and Bitter, Neal P and Koehler, Timothy P and Plimpton, Steven J and Papadakis, George},
  journal={Physical Review Fluids},
  volume={3},
  number={7},
  pages={071402},
  year={2018},
  publisher={APS}
}

@article{frapolli2015entropic,
  title={Entropic lattice Boltzmann model for compressible flows},
  author={Frapolli, Nicol{\`o} and Chikatamarla, Shyam S and Karlin, Iliya V},
  journal={Physical Review E},
  volume={92},
  number={6},
  pages={061301},
  year={2015},
  publisher={APS}
}

@article{atif2018higher,
  title={Higher-order lattice Boltzmann model for thermohydrodynamics},
  author={Atif, Mohammad and Namburi, Manjusha and Ansumali, Santosh},
  journal={Physical Review E},
  volume={98},
  number={5},
  pages={053311},
  year={2018},
  publisher={APS}
}

@article{namburi2016crystallographic,
  title={Crystallographic lattice Boltzmann method},
  author={Namburi, Manjusha and Krithivasan, Siddharth and Ansumali, Santosh},
  journal={Scientific reports},
  volume={6},
  pages={27172},
  year={2016},
  publisher={Nature Publishing Group}
}

@article{ansumali2007hydrodynamics,
  title={Hydrodynamics beyond Navier-Stokes: Exact solution to the lattice Boltzmann hierarchy},
  author={Ansumali, S and Karlin, IV and Arcidiacono, S and Abbas, A and Prasianakis, NI},
  journal={Physical review letters},
  volume={98},
  number={12},
  pages={124502},
  year={2007},
  publisher={APS}
}

@article{chen1998lattice,
  title={Lattice Boltzmann method for fluid flows},
  author={Chen, Shiyi and Doolen, Gary D},
  journal={Annual review of fluid mechanics},
  volume={30},
  number={1},
  pages={329--364},
  year={1998},
  publisher={Annual Reviews 4139 El Camino Way, PO Box 10139, Palo Alto, CA 94303-0139, USA}
}

@book{succi2001lattice,
  title={The lattice Boltzmann equation: for fluid dynamics and beyond},
  author={Succi, Sauro},
  year={2001},
  publisher={Oxford university press}
}

@article{benzi1992lattice,
  title={The lattice Boltzmann equation: theory and applications},
  author={Benzi, Roberto and Succi, Sauro and Vergassola, Massimo},
  journal={Physics Reports},
  volume={222},
  number={3},
  pages={145--197},
  year={1992},
  publisher={Elsevier}
}

@article{shan1998discretization,
  title={Discretization of the velocity space in the solution of the Boltzmann equation},
  author={Shan, Xiaowen and He, Xiaoyi},
  journal={Physical Review Letters},
  volume={80},
  number={1},
  pages={65},
  year={1998},
  publisher={APS}
}

@article{yudistiawan2010higher,
  title={Higher-order Galilean-invariant lattice Boltzmann model for microflows: Single-component gas},
  author={Yudistiawan, Wahyu Perdana and Kwak, Sang Kyu and Patil, DV and Ansumali, Santosh},
  journal={Physical Review E},
  volume={82},
  number={4},
  pages={046701},
  year={2010},
  publisher={APS}
}

@article{gallis2017molecular,
  title={Molecular-level simulations of turbulence and its decay},
  author={Gallis, MA and Bitter, NP and Koehler, TP and Torczynski, JR and Plimpton, SJ and Papadakis, G},
  journal={Physical Review Letters},
  volume={118},
  number={6},
  pages={064501},
  year={2017},
  publisher={APS}
}

@inproceedings{pradhan2016transition,
  title={Transition and turbulence in a wall-bounded channel flow at high Mach number},
  author={Pradhan, Sahadev and Kumaran, V},
  booktitle={52nd AIAA/SAE/ASEE Joint Propulsion Conference},
  pages={5050},
  year={2016}
}

@article{agrawal2018molecular,
  title={Molecular dice: Random number generators {\'a} la Boltzmann},
  author={Agrawal, Samarth and Bhattacharya, Soumyadeep and Ansumali, Santosh},
  journal={Physical Review E},
  volume={98},
  number={6},
  pages={063315},
  year={2018},
  publisher={APS}
}

@article{montessori2020multiresolution,
  title={A Multiresolution Mesoscale Approach for Microscale Hydrodynamics},
  author={Montessori, Andrea and Tiribocchi, Adriano and Lauricella, Marco and Bonaccorso, Fabio and Succi, Sauro},
  journal={Advanced Theory and Simulations},
  volume={3},
  number={4},
  pages={1900250},
  year={2020},
  publisher={Wiley Online Library}
}

@article{grad1949kinetic,
  title={On the kinetic theory of rarefied gases},
  author={Grad, Harold},
  journal={Communications on pure and applied mathematics},
  volume={2},
  number={4},
  pages={331--407},
  year={1949},
  publisher={Wiley Online Library}
}

@article{di2016dsmc,
  title={Dsmc--lbm mapping scheme for rarefied and non-rarefied gas flows},
  author={Di Staso, Gianluca and Clercx, HJH and Succi, Sauro and Toschi, Federico},
  journal={Journal of Computational Science},
  volume={17},
  pages={357--369},
  year={2016},
  publisher={Elsevier}
}

@article{kolluru2020lattice,
  title={Lattice Boltzmann model for weakly compressible flows},
  author={Kolluru, Praveen Kumar and Atif, Mohammad and Namburi, Manjusha and Ansumali, Santosh},
  journal={Physical Review E},
  volume={101},
  number={1},
  pages={013309},
  year={2020},
  publisher={APS}
}

@article{garcia1998generation,
  title={Generation of the Chapman--Enskog distribution},
  author={Garcia, Alejandro L and Alder, Berni J},
  journal={Journal of computational physics},
  volume={140},
  number={1},
  pages={66--70},
  year={1998},
  publisher={Academic Press}
}

@book{bird1994molecular,
  title={Molecular gas dynamics and the direct simulation of gas flows},
  author={Bird, Graeme A and Brady, JM},
  volume={5},
  year={1994},
  publisher={Clarendon press Oxford}
}

@article{tillmark1992experiments,
  title={Experiments on transition in plane Couette flow},
  author={Tillmark, Nils and Alfredsson, P Henrik},
  journal={Journal of Fluid Mechanics},
  volume={235},
  pages={89--102},
  year={1992},
  publisher={Cambridge University Press}
}

@article{lundbladh1991direct,
  title={Direct simulation of turbulent spots in plane Couette flow},
  author={Lundbladh, Anders and Johansson, Arne V},
  journal={Journal of Fluid Mechanics},
  volume={229},
  pages={499--516},
  year={1991},
  publisher={Cambridge University Press}
}

@article{schmiegel1997fractal,
  title={Fractal stability border in plane Couette flow},
  author={Schmiegel, Armin and Eckhardt, Bruno},
  journal={Physical review letters},
  volume={79},
  number={26},
  pages={5250},
  year={1997},
  publisher={APS}
}

@article{komminaho1996very,
  title={Very large structures in plane turbulent Couette flow},
  author={Komminaho, Jukka and Lundbladh, Anders and Johansson, Arne V},
  journal={Journal of Fluid Mechanics},
  volume={320},
  pages={259--285},
  year={1996},
  publisher={Cambridge University Press}
}

@article{hof2006finite,
  title={Finite lifetime of turbulence in shear flows},
  author={Hof, Bj{\"o}rn and Westerweel, Jerry and Schneider, Tobias M and Eckhardt, Bruno},
  journal={Nature},
  volume={443},
  number={7107},
  pages={59--62},
  year={2006},
  publisher={Nature Publishing Group}
}

@book{papanastasiou2021viscous,
  title={Viscous fluid flow},
  author={Papanastasiou, Tasos and Georgiou, Georgios and Alexandrou, Andreas N},
  year={2021},
  publisher={CRC press}
}

@book{leal2007advanced,
  title={Advanced transport phenomena: fluid mechanics and convective transport processes},
  author={Leal, L Gary},
  volume={7},
  year={2007},
  publisher={Cambridge University Press}
}

@article{aydin1979novel,
  title={Novel experimental facility for the study of plane Couette flow},
  author={Aydin, Mahir and Leutheusser, Hans J},
  journal={Review of Scientific Instruments},
  volume={50},
  number={11},
  pages={1362--1366},
  year={1979},
  publisher={American Institute of Physics}
}

@article{gallis2021turbulence,
  title={Turbulence at the edge of continuum},
  author={Gallis, MA and Torczynski, JR and Krygier, MC and Bitter, NP and Plimpton, SJ},
  journal={Physical Review Fluids},
  volume={6},
  number={1},
  pages={013401},
  year={2021},
  publisher={APS}
}

@article{wasow1953small,
  title={On small disturbances of plane Couette flow},
  author={Wasow, Wolfgang},
  journal={J. Res. Nat. Bur. Stand},
  volume={51},
  number={4},
  pages={195--202},
  year={1953}
}

@article{gallagher1962behaviour,
  title={On the behaviour of small disturbances in plane Couette flow},
  author={Gallagher, AP and Mercer, A McD},
  journal={Journal of Fluid Mechanics},
  volume={13},
  number={1},
  pages={91--100},
  year={1962},
  publisher={Cambridge University Press}
}

@article{davies1928experimental,
  title={An experimental study of the flow of water in pipes of rectangular section},
  author={Davies, SJ and White, CM},
  journal={Proceedings of the Royal Society of London. Series A, Containing Papers of a Mathematical and Physical Character},
  volume={119},
  number={781},
  pages={92--107},
  year={1928},
  publisher={The Royal Society London}
}

@article{kao1970experimental,
  title={Experimental investigations of the stability of channel flows. Part 1. Flow of a single liquid in a rectangular channel},
  author={Kao, Timothy W and Park, C},
  journal={Journal of Fluid Mechanics},
  volume={43},
  number={1},
  pages={145--164},
  year={1970},
  publisher={Cambridge University Press}
}

@article{patel1969some,
  title={Some observations on skin friction and velocity profiles in fully developed pipe and channel flows},
  author={Patel, VC and Head, MR},
  journal={Journal of Fluid Mechanics},
  volume={38},
  number={1},
  pages={181--201},
  year={1969},
  publisher={Cambridge University Press}
}

@article{chikatamarla2006grad,
  title={Grad's approximation for missing data in lattice Boltzmann simulations},
  author={Chikatamarla, SS and Ansumali, SKIV and Karlin, IV},
  journal={EPL (Europhysics Letters)},
  volume={74},
  number={2},
  pages={215},
  year={2006},
  publisher={IOP Publishing}
}

@book{succi2018lattice,
  title={The lattice Boltzmann equation: for complex states of flowing matter},
  author={Succi, Sauro and Succi, S},
  year={2018},
  publisher={Oxford University Press}
}

@article{chen2003extended,
  title={Extended Boltzmann kinetic equation for turbulent flows},
  author={Chen, Hudong and Kandasamy, Satheesh and Orszag, Steven and Shock, Rick and Succi, Sauro and Yakhot, Victor},
  journal={Science},
  volume={301},
  number={5633},
  pages={633--636},
  year={2003},
  publisher={American Association for the Advancement of Science}
}

@article{montessori2015lattice,
  title={Lattice Boltzmann approach for complex nonequilibrium flows},
  author={Montessori, A and Prestininzi, P and La Rocca, M and Succi, S},
  journal={Physical Review E},
  volume={92},
  number={4},
  pages={043308},
  year={2015},
  publisher={APS}
}

@article{benzi2006mesoscopic,
  title={Mesoscopic two-phase model for describing apparent slip in micro-channel flows},
  author={Benzi, R and Biferale, L and Sbragaglia, M and Succi, S and Toschi, F},
  journal={EPL (Europhysics Letters)},
  volume={74},
  number={4},
  pages={651},
  year={2006},
  publisher={IOP Publishing}
}

@article{lockerby2005usefulness,
  title={The usefulness of higher-order constitutive relations for describing the Knudsen layer},
  author={Lockerby, Duncan A and Reese, Jason M and Gallis, Michael A},
  journal={Physics of Fluids},
  volume={17},
  number={10},
  pages={100609},
  year={2005},
  publisher={American Institute of Physics}
}

@book{kogan2013rarefied,
  title={Rarefied gas dynamics},
  author={Kogan, Maurice N},
  year={2013},
  publisher={Springer}
}

@article{mcmullen2022navier,
  title={Navier-Stokes Equations Do Not Describe the Smallest Scales of Turbulence in Gases},
  author={Ryan M. McMullen, Michael C. Krygier, John R. Torczynski and Michael A. Gallis},
  journal={Physical Review Letters},
  volume={128},
  number={11},
  pages={114501},
  year={2022},
  publisher={APS}
}

\newpage

\section{Supplementary Material: \\ Fluid-kinetic multiscale solver for wall-bounded turbulence}


\textbf{Fluctuations in the near-wall region: }

\begin{figure}[H]
\centering
\begin{subfigure}[b]{0.43\textwidth}
    \includegraphics[scale=0.52]{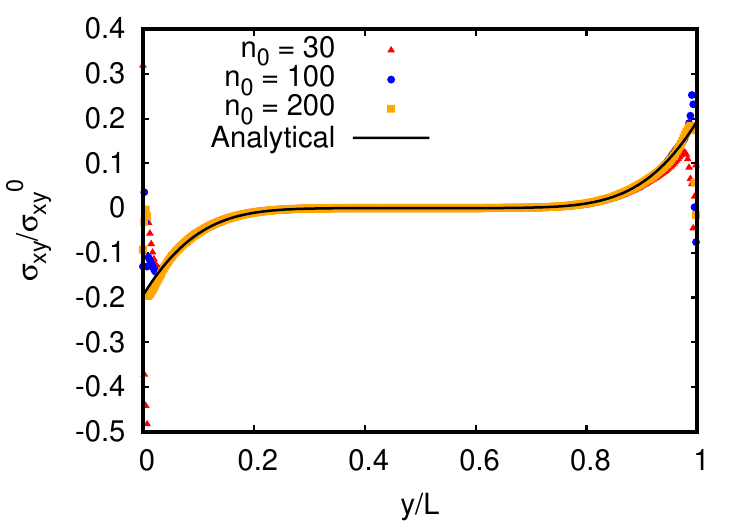}
    \caption{}
     \label{fig:shear_numPart}
\end{subfigure}
\hspace{10mm}
\begin{subfigure}[b]{0.43\textwidth}
    \includegraphics[scale=0.52]{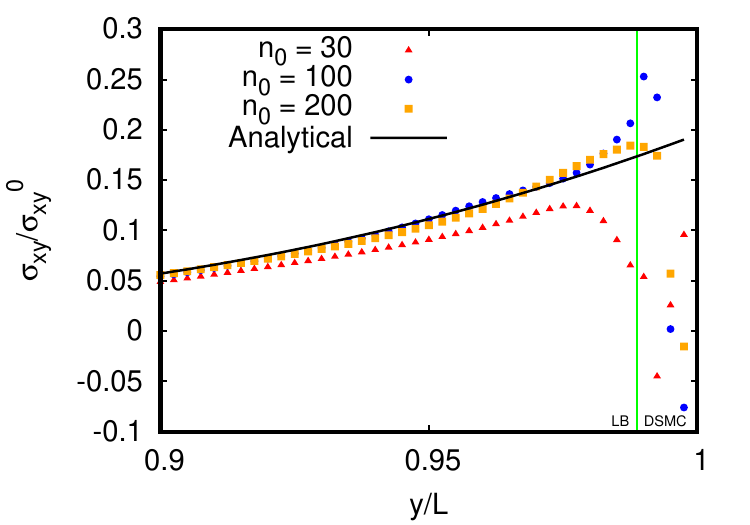}
    \caption{}
    \label{fig:shear_numPart_zoomed}
\end{subfigure}

\caption{(\protect\subref{fig:shear_numPart}) Normalized shear stress profile at $\bar{t} = 7.49*10^{-3}$ for simulations performed using 30, 100 and 200 particles per cell compared against the analytical solution. (\protect\subref{fig:shear_numPart_zoomed}) A magnified version of the near-wall region.}
\label{fig:shear_fluctuations}
\end{figure}

Simulations using 30, 100, and 200 particles per cell were performed in order to assess the effects of number density on the fluctuations in the near-wall stress profile.  This information is  reported in Figure \ref{fig:shear_fluctuations}\subref{fig:shear_numPart} \& \ref{fig:shear_fluctuations}\subref{fig:shear_numPart_zoomed}, from which it is 
seen that the standard deviation decreases with increasing $n_0$,  as also shown in Table \ref{tab:numPart_variance_shear}.

\begin{table}[H]
  \begin{center}
    \begin{tabular}{|c|c|}
	    \hline
      $n_0$ & $\sigma_{(\bar{y} \geq 0.9)}$ \\
      \hline
      $30 $ & $0.308959$\\
      $100$ & $0.176806$\\
      $200$ & $0.154661$\\
      \hline
    \end{tabular}
    \quad \quad
    \begin{tabular}{|c|c|}
	    \hline
      $n_0$ & $\sigma_{(\bar{y} \geq 0.95)}$ \\
      \hline
      $30 $ & $0.435186$\\
      $100$ & $0.249875$\\
      $200$ & $0.218377$\\
      \hline
    \end{tabular}
    \caption{Standard deviation ($\sigma$) of the shear stress fluctuations with different number of particles per cell, measured for data points $\bar{y} \geq 0.9$ and $\bar{y} \geq 0.95$. Here $\bar{y} = y/L$ is the nondimensional length.}
	\label{tab:numPart_variance_shear}
  \end{center}
\end{table}

\end{document}